\documentclass[twocolumn]{el-author}
\usepackage{tikz}
\usepackage{pgfplots}
\usepackage[nolist]{acronym}

\newcommand{\mat}[1]{\mathbf{#1}}
\newcommand{\T}{^\top}
\renewcommand{\H}{^\mathrm{H}}

\newcommand{\F}{\mathbf{F}}
\newcommand{\Z}{\mathbf{Z}}
\newcommand{\I}{\mathbf{I}}
\newcommand{\D}{\mathbf{D}}
\newcommand{\G}{\mathbf{G}}
\newcommand{\W}{\mathbf{W}}

\renewcommand{\O}{\mathcal{O}}
\newcommand{\ld}{\operatorname{ld}}

\date{\today}
\title{Reduced Complexity Calculation of LMMSE Filter Coefficients for GFDM}
\author{Maximilian Matthe, Ivan Gaspar, Dan Zhang, Gerhard Fettweis}

\newcommand{\CR}[1]{#1}

\abstract{
  A low-complexity algorithm for calculation of the LMMSE filter
  coefficients for GFDM in a block-fading multipath environment is
  derived in this letter. The simplification is based on the block
  circularity of the involved matrices.  The proposal reduces
  complexity from cubic to squared order.  \CR{The proposed approach
    can be generalized to other waveforms with circular pulse shaping.}
}

\begin{document}

\begin{acronym}
  \acro{CP}{cyclic prefix}
  \acro{DFT}{discrete Fourier transform}
  \acro{GFDM}{Generalized Frequency Division Multiplexing}
  \acro{BC}{block circulant}
  \acro{AWGN}{additive white Gaussian noise}
\end{acronym}

\maketitle
%
%
\section{Notation}
Matrices are typeset in bold notation. $\otimes$ denotes the Kronecker
product, $\mat{I}_N$ denotes the N-dimensional identity matrix and
$\mat{F}_N$ is the unitary N-point DFT matrix. $\left<x\right>_N$
denotes the remainder of $x$ modulo $N$. $(\cdot)\T$ and $(\cdot)\H$
denote matrix transpose and hermitian conjugate,
respectively. $\text{diag}(\dots)$ returns a (block)-diagonal
matrix with its arguments on the diagonal.

\section{Motivation and Problem Description}\label{sec-1}
\CR{Recently, several waveforms for 5G networks have been proposed
  \cite{Wunder2014}.  Among them, waveforms utilizing circular
  pulse shaping \cite{Lin2014,Tonello2014,Michailow2014} structure the signal into self-contained blocks that
  can be separated by a \ac{CP} to combat inter-block interference.}
\CR{\ac{GFDM}, the first 5G waveform that used circular pulse shaping, provides a flexible time
  and frequency grid that can be explored to provide low out of band radiation and robustness to time- and
frequency misalignments \cite{Michailow2014}}. For real-world implementations,
low-complexity algorithms are always of concern.  For \ac{GFDM},
literature provides low-complexity descriptions for the linear
\ac{GFDM} modulator and demodulator \cite{Michailow2014} and the
design of zero-forcing and LMMSE filters for \ac{AWGN} channels
\cite{Matthe2014}.  However, no low-complexity implementation of the
LMMSE demodulator for multipath environments is available.  This
letter presents an algorithm with significantly reduced complexity for
calculating the LMMSE filter coefficients for \ac{GFDM}.

The time domain signal of one \ac{GFDM} block is given by
\begin{align}\label{eq:tx}
  x[n] &= \sum_{k=0}^{K-1}\sum_{m=0}^{M-1}d_{k,m} g_{k,m}[n]\\
\text{with }  g_{k,m}[n] &= g[\left<n-mK\right>_N]\exp\left(j2\pi\frac{kM}{N}n\right),
\end{align}
where $N=KM$, $n=0,\dots,N-1$, $g[n]$ denotes the prototype transmit filter and $d_{k,m}\in\mathcal{X}$ is the data symbol to be transmitted on the $k$th subcarrier and $m$th subsymbol taken from the complex-valued constellation $\mathcal{X}$.  Eq. (\ref{eq:tx}) is written in matrix form as $\vec{x}= \mat{A}\vec{d}$ with
\begin{align}\label{eq:Amat}
  \mat{A}&=\left[\vec{g}_{0,0}, \vec{g}_{1,0}, \dots, \vec{g}_{K-1,0}, \vec{g}_{0,1},\vec{g}_{1,1}
  \dots \vec{g}_{K-1,M-1}\right],
\end{align}
where the column vectors are $\vec{g}_{k,m}=(g_{k,m}[n])_{n=0,\dots,N-1}$, $\vec{d}$ contains $d_{k,m}$ in the appropriate order and $E[\vec{d}]=\mat{0}$, $E[\vec{d}\vec{d}\H]=\mat{I}_N$.  $g[n]$ is bandlimited within two subcarriers, i.e. $\F_N\vec{g}_{k,0}$ only has $2M$ nonzero elements centered around the index $kM$ \cite{Michailow2014}. The signal is transmitted through a block-fading wireless multipath channel with impulse response $\vec{h}$. Assuming a \ac{CP} between blocks that is longer than $\vec{h}$, the received time domain signal per block is given by
\begin{align}\label{eq:rx}
  \vec{y}&=\mat{H}\vec{x}=\mat{H}\mat{A}\vec{d}+\vec{w},
\end{align}
where $\mat{H}$ is an $N\times N$ circulant matrix with the zero-padded channel impulse response $\vec{h}$ as its first column and \mbox{$\vec{w} \sim \mathcal{CN}(0, \sigma_{n}^2\mat{I}_N)$} is \ac{AWGN}. From (\ref{eq:rx}), the LMMSE equalizer for $\vec{d}$ is given by \(\vec{d}_{\text{LMMSE}}=\mat{W}\H \vec{y} \),
where $\mat{W}$ are the LMMSE filter coefficients given by
\begin{align}\label{eq:wlin}
  \mat{W}&=\mat{H}\mat{A}((\mat{H}\mat{A})\H(\mat{H}\mat{A})+\sigma_n^2\mat{I}_N)^{-1}.
\end{align}
Sec II. reviews properties of \ac{BC} matrices. A low-complexity solution for computing $\mat{W}$ in (\ref{eq:wlin}) is developed in Sec. III and its complexity is evaluated in Sec IV. Sec V concludes this letter.

\section{Block-circulant Matrices}
Let $\mat{X}$ be an $N\times N$ \ac{BC} matrix composed of $M$ arbitrary submatrices $\{\mat{X}_{m}\}$ of size $K\times K$ each, i.e.
\begin{align}
  \mat{X}&=
  \begin{pmatrix}
    \mat{X}_0 & \mat{X}_{M-1} & \mat{X}_{M-2}&\dots & \mat{X}_{1}\\
    \mat{X}_1 & \mat{X}_{0}&\mat{X}_{M-1}\\
    \mat{X}_2 & \mat{X}_{1} & \mat{X}_{0}& \\
    \vdots &&&\ddots&\vdots\\
    \mat{X}_{M-1} & \mat{X}_{M-2} & \dots& &\mat{X}_{0}
  \end{pmatrix}.
\end{align}
$\mat{X}$ is block-diagonalized by $\mat{Z}=\mat{F}_M\otimes \mat{I}_K$ \cite{Qiu1995} such that
\begin{align}
  \Z\mat{X}\Z\H&=\text{diag}(\D_{\mat{X},0},\D_{\mat{X},1},\dots,\D_{\mat{X},M-1}),
\end{align}
where $\D_{X,u}$ is the $u$th submatrix of size $K\times K$ on the diagonal of $\Z\mat{X}\Z\H$.  Note that $\Z$ performs a discrete ZAK transform on its argument \cite{Matthe2014}.  Let $\mat{X}_s$ be the first $K$ colums of $\mat{X}$, i.e. $\mat{X}_s=\begin{pmatrix}\mat{X}_0\T & \mat{X}_{1}\T&\cdots&\mat{X}_{M-1}\T\end{pmatrix}\T$. Then \cite{Qiu1995},
\begin{align}
  \D_{\mat{X}}=
  \begin{pmatrix}
  \D_{\mat{X},0}\T & \D_{\mat{X},1}\T & \dots \D_{\mat{X},M-1}\T
  \end{pmatrix}\T&=\Z \mat{X}_s.
\end{align}
Note the similar behaviour of circulant matrices: Let $\mat{Y}$ be a circulant matrix with $\vec{y}$ in the first column, then \mbox{$\F_N\mat{Y}\F_N\H=\text{diag}(\F_N\vec{y})$}. Let\vspace{-2mm}\\
\begin{align}\Z_{u}&=\vec{\omega}_u\otimes \I_K\\\text{with }\vec{\omega}_{u}&=\begin{pmatrix}
  1 & \omega^{u} & \omega^{2u} &\cdots & \omega^{(M-1)\cdot u}
\end{pmatrix}
\end{align}
and $\omega=\exp(\tfrac{-j2\pi}{M})$ such that $\D_{\mat{X},u}=\Z_u\mat{X}_s$. Note that, according to above block diagonalization, the product and sum of two or the inverse of one \ac{BC} matrix is again \ac{BC}.

\section{Reduced complexity Filter Calculation}
From definition (\ref{eq:Amat}), $\mat{A}$ is \ac{BC} with $M$ blocks with size $K\times K$ and is hence block-diagonalized by $\Z$. Also, circularity of $\mat{H}$ implies block circularity. Accordingly, $\mat{G}=\mat{H}\mat{A}$, $(\G\H\G+\sigma_n^2\I_N)$ and $\W$ are all \ac{BC} matrices.  Hence, $\W$ is completely defined by its diagonalization $\D_{\W}=\Z\W_s$. The $M$ blocks of $\D_{\W}$ are given by the M equation systems
\begin{align}\label{eq:wBlock}
  \D_{\W,u}&=((\D_{\G,u}\H\D_{\G,u}+\sigma_n^2\I_K)^{-1}\D_{\G,u}\H)\H
\end{align}
where $u=0,\dots,M-1$ and $\D_{\G,u}=\Z_u\G_{s}$.  Now, $\Z_{u}\H\Z_u=((\vec{\omega}_u)\H\vec{\omega}_u)\otimes \I_K$ is a circulant matrix since \mbox{$((\vec{\omega}_u)\H\vec{\omega}_{u})_{i,j}=\omega^{u(i-j)}$} is circulant and accordingly $\D_{u}:=\F_N\Z_u\H\Z_u\F_N\H$ is a diagonal matrix. Due to band-limitation of $g[n]$ only adjacent subcarriers overlap and $(\F_N\G_s)\H \D_{u} \F_N\G_s$ is a tridiagonal matrix with periodic boundary conditions.
Once $\D_{\W}$ is known, the first K columns of $\W$ in the time
domain are given by
\begin{align}
  \W_s&=\Z\H\D_{\W},
\end{align}
and remaining columns are given as circular shifts of $\W_s$.  \CR{In
  addition, the ZAK domain $\D_{\W}$ can be also directly transformed
  into the frequency domain to readily employ a low-complexity
  receiver as in \cite{Gaspar2013}. Furthermore, LMMSE filtering can
  even be directly performed in the ZAK domain by
  \begin{align}
    \Z\vec{d}_{\text{LMMSE}}&=\text{diag}(\D_{\W,0},\D_{\W,1},\dots,\D_{\W,M-1})\H \Z \vec{y}
  \end{align}
}

\section{Complexity Analysis}
In this section, the arithmetic complexity of the proposed algorithm
is evaluated, considering one complex multiplication as one
$\mathcal{O}(1)$ operation and neglecting other operations such as
additions.  The product $\Z \vec{x}$ is equivalent to $K$ DFTs of
length $M$ each, and hence requires $K M\ld M$\footnote{Assuming that
  the $N$-point DFT requires $N\ld N$ operations.} operations,
yielding $K^2 M\ld M$ operations for $\Z \mat{X}_s$.

To compute $\F_N\G_s=\F_N\mat{H}\mat{A}_s$ we take advantage of the factorization
\begin{align}
  \F_N\G_s&=\F_N\mat{H}\F_N\H\F_N\mat{A}_s
\end{align}
where $\F_N\mat{A}_s$ is precalculated at the receiver. This is done by DFT for the first column. Other columns are given by circular shifts since $\mat{A}_s$ contains frequency shifts of $\vec{g}$. Considering band-limitation of $g[n]$, the multiplication of $\F_N\mat{A}_{s}$ with the diagonal matrix $\F_N\mat{H}\F_N\H$ takes $2MK$ operations. The diagonal of $\F_N\mat{H}\F_N\H$ is assumed to be available from previous channel estimation procedures. The product $\G_s\H\F_{N}\H\D_{u}\F_N\G_s$ requires $3K\cdot 4M$ operations due to the tridiagonal structure of the result and band-structure of $\F_N\G_{s}$.

The tridiagonal system with periodic boundary conditions
\begin{align}
  (\D_{\G,u}\H\D_{\G,u}+\sigma_n^2\I_K)^{-1}\D_{\G,u}
\end{align}
is solved \CR{using the Thomas algorithm \cite{Thomas}} with $2K$
operations for factorization and $5K$ operations for solving for each
right hand side, resulting in $2K+5K^2$ operations for the full linear
system \cite{Golub1996}.  Finally, calculation of $\Z\H\D_{\W}$
requires $K^2 M\ld M$ operations.

Hence, the number of complex multiplications $C_{\text{sparse}}$ to solve (\ref{eq:wlin}) with the proposed method is given by
\begin{align}
  C_{\text{sparse}} &= \underbrace{2MK}_{(a)}+M(\underbrace{3K\cdot4M}_{(b)}+\underbrace{2K+5K^2}_{(c)})+\underbrace{K^2M\ld M}_{(d)}\\
&=K^2(5M+M\ld M)+K(12M^2+4M)\\&=\O(K^2M\ld M+KM^2)
\end{align}
where (a) corresponds to calculation of $\F_N\G_s$, (b) respects
$\G_s\H\F_N\H \D_u\F_N\G_S$, (c) describes the solution of the
tridiagonal system and (d) accounts for $\Z\H\D_{W}$.  For comparison,
direct application of a conventional Hermitian positive definite
solver to (\ref{eq:wlin}) requires
$C_{\text{direct}}=\frac{N^3}{3}+N\cdot2N^2$ operations only for the
solution step\footnote{i.e. product of $\mat{HA}$ etc. is not
  considered}, where the first term corresponds to Cholesky
decomposition and the second term refers to backward and forward
substitution for $N$ right-hand sides.  \CR{An additional advantage is
the reduced memory requirement of the proposed algorithm, as it
suffices to store the $MK^2$ filter coefficients for $\W_s$ instead of
$M^2K^2$ coefficients for $\W$.}

\begin{figure}
  \centering
  \includegraphics[width=3.3in]{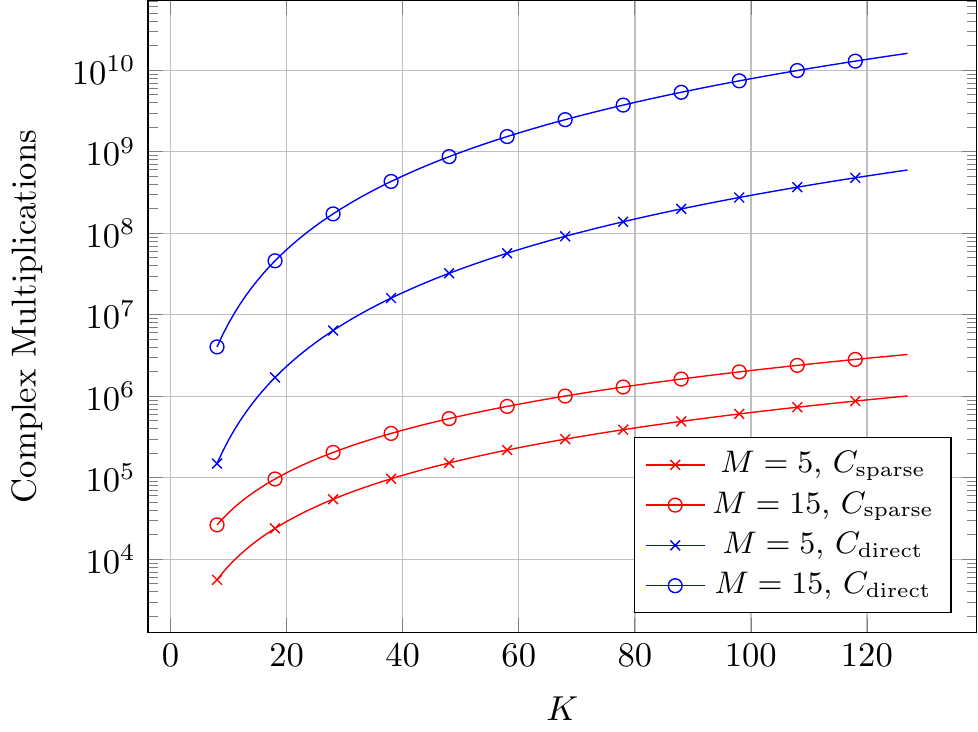}
  \caption{Number of floating point operations for proposed low-complexity LMMSE calculation.}
\label{fig:comp}
\end{figure}

Fig. \ref{fig:comp} compares the number of complex multiplications required for the proposed technique and for conventional solving with Cholesky decomposition for different values of $K$ and $M$.  The number of required operations can be reduced by 4 orders of magnitude for 128 subcarriers.

\section{Conclusion}
A low-complexity approach for the calculation of LMMSE filter
coefficients for block-fading multipath channels for GFDM has been
presented. The proposal significantly reduces the complexity of the
design from $\O(K^3M^3)$ to \mbox{$\O(K^2M\ld M+KM^2)$} which results
in a complexity reduction of several orders of magnitude for
reasonable system sizes.  \CR{Since the technique exploits the
block-circulant structure of the modulation matrix, it can be
generalized to other multicarrier waveforms employing circular pulse
shaping.}

\section{Acknowledgement}
This work has been performed in the framework of
ICT-619555 ``RESCUE'' and ICT-318555 ``5GNOW''
which are partly funded by the European Union.
\vskip5pt
\noindent Maximilian Matthe, Ivan Gaspar, Dan Zhang, Gerhard Fettweis
(\textit{Technical University Dresden, Vodafone Chair Mobile
  Communication Systems, Germany})
\vskip5pt
\noindent Email: \texttt{firstname.lastname@ifn.et.tu-dresden.de}

\vspace{5mm}
\bibliographystyle{IEEEtran}
\bibliography{library}

\begin{thebibliography}{1}
\providecommand{\url}[1]{#1}
\csname url@samestyle\endcsname
\providecommand{\newblock}{\relax}
\providecommand{\bibinfo}[2]{#2}
\providecommand{\BIBentrySTDinterwordspacing}{\spaceskip=0pt\relax}
\providecommand{\BIBentryALTinterwordstretchfactor}{4}
\providecommand{\BIBentryALTinterwordspacing}{\spaceskip=\fontdimen2\font plus
\BIBentryALTinterwordstretchfactor\fontdimen3\font minus
  \fontdimen4\font\relax}
\providecommand{\BIBforeignlanguage}[2]{{%
\expandafter\ifx\csname l@#1\endcsname\relax
\typeout{** WARNING: IEEEtran.bst: No hyphenation pattern has been}%
\typeout{** loaded for the language `#1'. Using the pattern for}%
\typeout{** the default language instead.}%
\else
\language=\csname l@#1\endcsname
\fi
#2}}
\providecommand{\BIBdecl}{\relax}
\BIBdecl

\bibitem{Wunder2014}
G.~Wunder, P.~Jung, M.~Kasparick, T.~Wild, F.~Schaich, Y.~Chen, S.~Brink,
  I.~Gaspar, N.~Michailow, A.~Festag, L.~Mendes, N.~Cassiau, D.~Ktenas,
  M.~Dryjanski, S.~Pietrzyk, B.~Eged, P.~Vago, and F.~Wiedmann,
  ``\BIBforeignlanguage{English}{{5GNOW: non-orthogonal, asynchronous waveforms
  for future mobile applications}},'' \emph{\BIBforeignlanguage{English}{IEEE
  Communications Magazine}}, vol.~52, no.~2, pp. 97--105, Feb. 2014.

\bibitem{Lin2014}
\BIBentryALTinterwordspacing
H.~Lin and P.~Siohan, ``{Multi-carrier modulation analysis and WCP-COQAM
  proposal},'' \emph{EURASIP Journal on Advances in Signal Processing}, vol.
  2014, no.~1, p.~79, 2014. [Online]. Available:
  \url{http://asp.eurasipjournals.com/content/2014/1/79}
\BIBentrySTDinterwordspacing

\bibitem{Tonello2014}
\BIBentryALTinterwordspacing
A.~M. Tonello and M.~Girotto, ``{Cyclic block filtered multitone modulation},''
  \emph{EURASIP Journal on Advances in Signal Processing}, vol. 2014, no.~1, p.
  109, 2014. [Online]. Available:
  \url{http://asp.eurasipjournals.com/content/2014/1/109}
\BIBentrySTDinterwordspacing

\bibitem{Michailow2014}
N.~Michailow, M.~Matth\'{e}, I.~Gaspar, A.~{Navarro Caldevilla}, L.~L. Mendes,
  A.~Festag, and G.~Fettweis, ``{Generalized Frequency Division Multiplexing
  for 5th Generation Cellular Networks},'' \emph{IEEE Transactions on
  Communications}, vol.~62, no.~9, pp. 3045--3061, 2014.

\bibitem{Matthe2014}
M.~Matth\'{e}, L.~L. Mendes, and G.~Fettweis, ``{GFDM in a Gabor Transform
  Setting},'' \emph{IEEE Communications Letters}, vol.~18, no.~8, pp.
  1379--1382, 2014.

\bibitem{Qiu1995}
\BIBentryALTinterwordspacing
S.~Qiu, ``{Block-circulant Gabor-matrix structure and discrete Gabor
  transforms},'' \emph{Optical Engineering}, vol.~34, no.~10, p. 2872, Oct.
  1995. [Online]. Available:
  \url{http://opticalengineering.spiedigitallibrary.org/article.aspx?articleid=1073919}
\BIBentrySTDinterwordspacing

\bibitem{Gaspar2013}
I.~S. Gaspar, M.~N., A.~{Navarro Caldevilla}, E.~Ohlmer, S.~Krone, and
  G.~Fettweis, ``{Low Complexity GFDM Receiver Based On Sparse Frequency Domain
  Processing},'' in \emph{Vehicular Technology Conference, 2013. VTC Spring
  2013, IEEE 77th}, 2013.

\bibitem{Thomas}
L.~H. Thomas, ``{Elliptic problems in linear difference equations over a
  network},'' Columbia University, Tech. Rep., 1949.

\bibitem{Golub1996}
\BIBentryALTinterwordspacing
G.~H. Golub and C.~F.~V. Loan, \emph{{Matrix Computations}}, 1996. [Online].
  Available:
  \url{http://books.google.de/books/about/Matrix\_Computations.html?id=mlOa7wPX6OYC\&pgis=1}
\BIBentrySTDinterwordspacing

\end{thebibliography}

\end{document}